\documentclass[a4paper,10pt,twoside]{cpc-hepnp}
\usepackage{multicol}
\usepackage{graphicx}
\usepackage{ctable}
\usepackage{booktabs}
\usepackage{amssymb,bm,mathrsfs,bbm,amscd}
\usepackage[tbtags]{amsmath}
\usepackage{lastpage}
\usepackage{hyperref}

\begin{document}

\title{Experimental Search for Dark Matter in China}
\pagestyle{plain}
\thispagestyle{fancy}          
\rhead{Submitted to `Frontiers of Physics'}

\author{Li Zhao$^{1,*}$, Jianglai Liu$^{1,2,\dag}$}

\maketitle
\address{
$^1$School of Physics and Astronomy, Shanghai Jiao Tong University, Shanghai 200240, China\\
  $^2$Tsung-Dao Lee Institute, Shanghai 200240, China\\
  Corresponding authors.\ E-mail: $^*$zhaoli78@sjtu.edu.cn,
  $^\dag$jianglai.liu@sjtu.edu.cn \\}

\begin{abstract}
The nature of dark matter is one of the greatest mysteries in
  modern physics and astronomy. A wide variety of experiments have
  been carried out worldwide to search for the evidence of particle
  dark matter. Chinese physicists started experimental search
  for dark matter about ten years ago, and have produced results with
  high scientific impact. In this paper, we present an overview of
  the dark matter program in China, and discuss recent results and
  future directions.
\end{abstract}



\tableofcontents

\section{Introduction}

\noindent
The original concept of galactic invisible matter, the dark matter,
has existed for almost a century, traditionally credited to Fritz
Zwicky~\cite{F-Zwicky}. Convincing observations came around in 1970s
from Vera Rubin and Kent Ford~\cite{Rubin-Ford} on the rotation curves
of many different galaxies, which strongly indicated the dominance of
galactic dark matter over the normal stars and gases. To date, the
gravitational effects of dark matter is evident from galaxies to
galaxy clusters and superclusters (see, for example,
Ref.~\cite{Steven}). During the past two decades, a standard model of
the cosmology, the so-called $\Lambda$CDM model, has
emerged~\cite{Kolb-and-Turner-Book}. This simple theoretical paradigm
gives remarkable agreements to a diverse set of astrophysical data from
the early universe (cosmic microwave background), large structure of
the universe, galactic dynamics, etc.  Precision cosmology studies
using the data, e.g. by the Planck satellite~\cite{Planck}, reveal
that the universe is composed of $4.9\%$ of baryonic matter, $26.8\%$
of cold dark matter, and $68.3\%$ of dark energy.

Microscopically, all known matters are made out of elementary
particles, described elegantly by the Standard Model of particle
physics~\cite{SM}. The latest triumph of the theory is the discovery
of the Higgs particle, the last missing particle in the theory, in
2012~\cite{ATLAS, CMS}. However, the Standard Model does not have any
viable dark matter candidates. Many classes of theories on the physics
beyond the Standard Model have been proposed in the past a few
decades, for example, Supersymmetry~\cite{SUSY}, Extra
Dimensions~\cite{EXD1,EXD2}, etc. These theories naturally predict
heavy, neutral, and stable particles, which can be the particle dark
matter candidates~\cite{SUSY-DM, EXD-DM}. Most of these theories also
predict a feeble non-gravitational interactions among the dark matter
particles and between the dark matter and normal matter. Generically,
these dark matter particles are referred to as the Weakly Interacting
Massive Particles (WIMPs), with a typical mass range between tens of
GeV/$c^2$ to a few TeV/$c^2$.

The WIMP paradigm fits elegantly with the thermal history of the
universe in the $\Lambda$CDM~\cite{Kolb-and-Turner-Book}. In the very
early hot universe, WIMPs were in thermal equilibrium with ordinary
matters, with a balance between the annihilation of WIMPs into
ordinary particle-antiparticle pairs and backwards. Then as the
universe expanded and cooled down, the annihilation reaction rates
fell below the expansion, so WIMPs became a thermal
relic. Coincidentally, most WIMP theories can reproduce the observed
relic density of the dark matter - this so-called ``WIMP Miracle''
offers very compelling motivation for this class of models.

The feeble interactions between the WIMPs and normal matters also
allow experimental studies in laboratories.  There are three
complementary approaches to identify WIMP dark matters. The first is
the so-called direct detection, in which galactic WIMPs can interact
with atomic nuclei in the target, resulting in nuclear recoils
(NRs). Background gamma rays will result in electron recoils (ERs),
which may be rejected experimentally. The second is the so-called
indirect detections, to search for high energy particles produced by
WIMP annihilations (e.g. from the center of the Galaxy). The third
type is to produce ``man-made" dark matter in high energy particle
colliders. All three directions are pursued intensively worldwide. In
China, experimental programs on dark matter direct and indirect
detections have gained significant momentum in the past ten years. We
shall present a review in these two areas in the rest of this paper.

\section{Direct detection experiments underground}
\noindent Due to the gravitational attractions, our Galaxy is
surrounded by a large dark matter halo, extended far beyond the range
of visible stars. The spacial distribution of the halo can be derived
from the dynamics of our Galaxy. Close to the solar orbit, the nominal
value for the dark matter mass density $\rho_{0}$ is
$0.3\pm0.1$~GeV/cm$^{3}$~\cite{SmithMC, SavageC}.  The cold dark
matter conjecture implies that these particles are ideal-gas-like with
no preferred direction of motion except being gravitationally
bound. Their velocity distribution is a Maxwell-Boltzmann distribution
with a velocity cutoff at 650 km/s, corresponding to the galactic
escape velocity~\cite{SmithMC, SavageC}.

The direct detection of the dark matter utilizes the fact that the
solar system orbits the Galactic center (and the dark matter halo on
average) with a nearly circular velocity
$\upsilon_{0}=220$~km/s~\cite{SmithMC, SavageC}. Therefore, relatively
to the terrestrial detectors, the incoming dark matter particles carry
an average momentum. The scattering of these dark matter particles
with known matters in a standard target can be approximated as a
non-relativistic two-body scattering. Direct detection experiments are
designed to observe the low-energy ($\sim$10~keV) recoils of the
nucleus struck rarely by an incoming WIMP. In principle, the electrons
can also interact with the WIMPs, but the recoil energy is
kinematically suppressed.

The recoil energy will be deposited in the target and converted into
excitations such as photons, ionization electrons, or phonons, and can
be detected by sensitive detectors. For example, one class of
experiments operating at very low temperature (below 100 mK, also
known as the micro-bolometer) is designed to detect the
phonons. Representative experiments are CDMS (Si~\cite{CDMS-Si} and
Ge~\cite{CDMS-Ge}) and CRESST-II~\cite{CRESST-CaWO4}(CaWO$_4$), which,
at the same time, detect the ionization and phonon signals,
respectively.  Another class of low noise semi-conductor experiments,
for example CoGeNT~\cite{CoGeNT2013} and CDEX~\cite{CDEX}, detects the
ioniziation signals only.  These experiments are sensitive to low mass
WIMPs (10 GeV/$c^2$ and below), which produce relatively small recoil
energy.  On the other hand, noble liquid detectors such as
XENON~\cite{XENON100-2012}, LUX~\cite{LUX-2014},
PandaX~\cite{PandaX} (xenon), DarkSide~\cite{DarkSide-2013} and
DEAP~\cite{DEAP-2008} (argon) typically detect scintillation photons,
some in conjunction with ionization electrons. These experiments are
powerful for the high mass WIMPs at the typical electroweak symmetry
breaking scale (100~GeV/$c^2$ and above).

\begin{figure}[!htbp]
\centering
\includegraphics[width=0.5\textwidth]{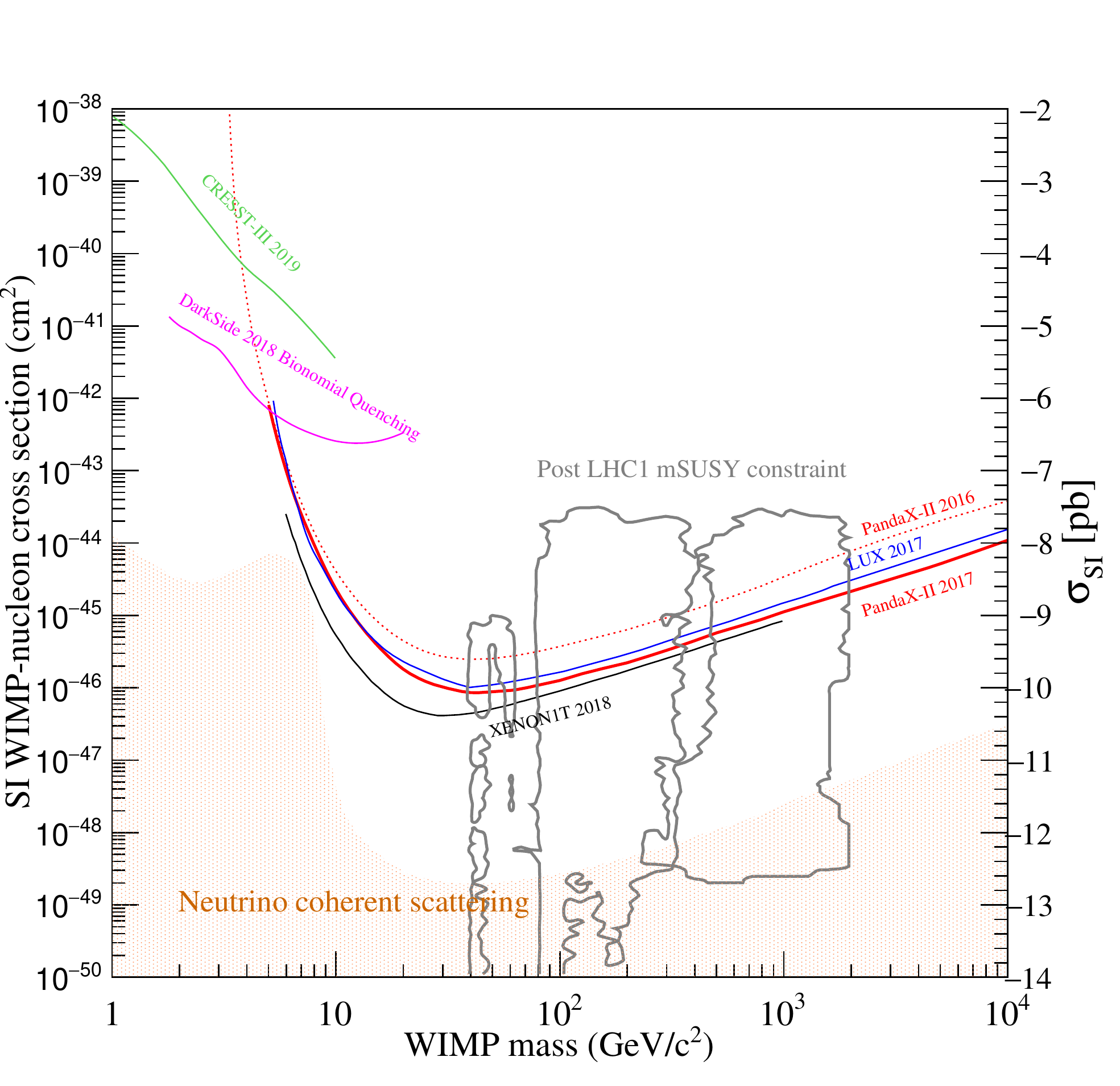}
\caption{The current 90\% upper limits on WIMP-nucleon SI
  cross-section. See legend for the leading exclusion limits from
  PandaX-II~\cite{PandaX-II-33ton-day,PandaX-II-50ton-day},
  XENON1T~\cite{XENON2018}, DarkSide~\cite{DarkSide50-2018}, and
  CRESST-III~\cite{CRESST2019}. The dotted region at the lower left
  indicates the ``limiting floor'' to the sensitivity below which one
  could not identify dark matter from the coherent nuclear recoil
  background from the solar and atmospheric neutrinos.
  \protect\label{cross-section} }
\end{figure}

For a given WIMP mass and the number of nuclear targets in a detector,
the detected rate (limit) can be converted to a measurement (limit) of
the scattering cross section between the WIMP and the nucleon. Recent
world results are shown in Fig.~\ref{cross-section}. Despite very
active search worldwide, none of the experiment has observed a
convincing signal yet. Ultimately, the sensitivity of the direct
detection experiments will be limited by the irreducible background
due to coherent nuclear scattering of solar and atmospheric neutrinos
- this is the so-called ``neutrino floor''~\cite{Nu-Floor}.

Due to the rarity of the dark matter signals, it is crucial to
maintain an ultralow background environment. All dark matter direct
detection experiments operate deep underground to reduce background
events from the cosmic rays. The development of the underground
laboratory provided a strong boost to the domestic
dark matter experimental program. The initial development started in
2009, jointly by the Yalong River Hydropower Development Company and
Tsinghua University, utilizing the 17.5 km horizontal access tunnels
through the Jinping Mountain in Sichuan, China\cite{CJPL}. With a rock
overburden of about 2400 meters, it is the deepest operating
underground laboratory in the world. The muon flux is measured to be
about 1/m$^{2}$/week~\cite{moun-flux-cjpl}, which minimizes the
cosmogenically induced background. The horizontal tunnel facilitates
the access.

The CJPL Phase-I (CJPL-I) lab consists of an experimental hall with an
approximate dimension of 6~m(W)$\times$ 6~m(H)$\times$40~m(L). The
space is currently occupied by CDEX~\cite{CDEX} and
PandaX~\cite{PandaX}, the first generation dark matter direct
detection experiments in China. A general purpose low radiopurity
screening facility with a few high purity Germanium detectors is also
operating in CJPL-I.

In 2015, much more experimental space has been excavated about 1500~m
away from CJPL-I. This new space is now called CJPL-II, with eight
experimental halls, each with dimensions of
14~m(W)$\times$14~m(H)$\times$65~m(L)~\cite{CJPL-2017}. China has
committed a major development project to make CJPL-II a general
purpose, deep underground, ultralow background national
facility. Future projects in dark matter, neutrinos, nuclear
astrophysics, geomechanics, etc., are being considered by the
scientific committee of the laboratory.

\subsection{PandaX dark matter experiment}
\begin{figure}[!htbp]
  \centering
\includegraphics[width=0.5\textwidth]{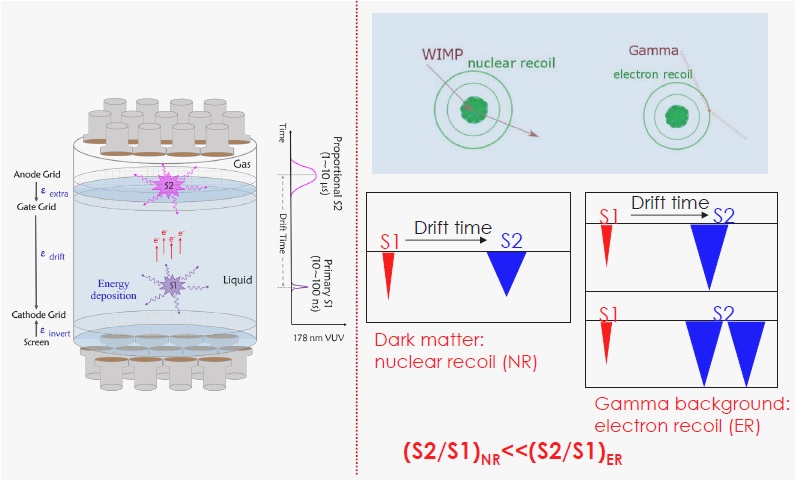}
\caption{The schematics of a two-phase xenon time projection
  chamber. Also shown are the illustrative signals for a gamma and
  WIMP event that cause either an electron or nuclear
  recoil. \protect\label{fig-TPC}}
\end{figure}

\noindent
The PandaX (Particle AND Astrophysical Xenon observatory) is a staged
experimental program aiming to use xenon as the target to search for
WIMP dark matter and to study the nature of neutrinos. The first
(PandaX-I, 120 kg) and second stage (PandaX-II, 580 kg) experiments
were completed in 2014 and 2019, respectively. Both experiments were
located in CJPL-I. PandaX-I and PandaX-II utilized the so-called
two-phase xenon time projection chamber (TPC) technique to detect the
energy and position of the WIMP-nucleus scattering. The operation
principle is shown in Fig.~\ref{fig-TPC}. A cryostat is filled with
liquid xenon, with two closely-packed arrays of photomultiplier (PMTs)
located at the top (gas) and bottom (liquid) of the chamber. The
prompt scintillation photons collected by the PMTs is referred as the
$S1$ signal. To collect the ionization electrons, a drift electric
field is applied between a light transmitting cathode and gate
electrodes, located at bottom and top of the liquid xenon,
respectively. A much stronger electrical field is set between the gate
(in liquid) and anode (in gas). This so-called extraction field
extracts the ionized electrons into the gaseous xenon, in which region
secondary electroluminescence photons are subsequently produced close
to the top PMT array. This delayed flash of photons is referred to as
the $S2$ signal. A three-dimensional imaging of the interaction vertex
can be achieved by using the pattern of $S2$ collected by the top PMT
array (horizontal position) and the time separation between the $S1$
and $S2$ (vertical position). Peripheral background coming from
outside of the detector can be significantly suppressed by
fiducialization. Events with multiple $S2$s (Fig.\ref{fig-TPC}) from
multiscattering of the background gammas or neutrons are also
rejected. For the remaining single-scatter events, due to the
difference in the ionization capability of the recoiling electron
(stronger) and nucleus (weaker), the fraction of energy carried by
ionized electrons is different between the two. Therefore the ratio
$S2/S1$ is another powerful discriminant against the ER background.
The superior potential of the liquid xenon TPC had been demonstrated
by several experiments, including XENON10~\cite{XENON10-2008},
XENON100~\cite{XENON100-2012}, ZEPLIN~\cite{ZEPLIN-II-2007}, and
LUX~\cite{LUX-2014}.

At CJPL-I, a passive shielding to suppress ambient neutrons and gamma
rays was constructed and used in both PandaX-I and PandaX-II. From
outside to inside, it consisted of 40~cm of Polyethylene (PE), 20~cm
of lead, 20~cm of PE, and 5~cm oxygen-free high thermal conductivity
copper (OFHC). The innermost shield is a vacuum copper vessel, which
is also a vacuum jacket for the cryostat and an ambient radon barrier.

\begin{figure}[!htbp]
\centering
\includegraphics[width=0.5\textwidth]{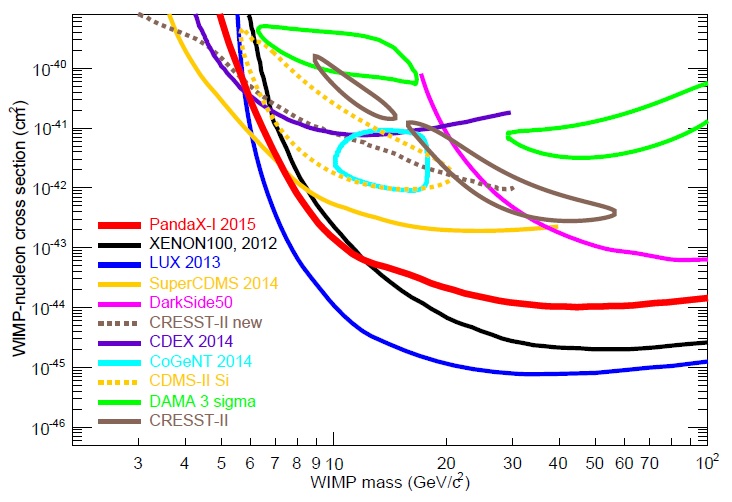}
\caption{The $90\%$ confidence level upper limit for SI isoscalar
  WIMP-nucleon cross section for the PandaX-I experiment (red
  curve). Results from other experiments are overlaid for comparison
  (see legend, contours are claimed signal regions, lines are
  exclusion limits).  Figure from Ref.~\cite{PandaX-I-2014}.
  \protect\label{fig-resultpandax1}}
\end{figure}

The PandaX-I TPC was specially designed for a high light yield thus
low energy threshold, to search for low-mass WIMPs. The TPC field cage
was enclosed by a cylindrical polytetrafluoroethylene (PTFE) wall with
15~cm in height and 60~cm in diameter. The pancake-shaped target was
to reduce photon loss before it reached the PMTs. For the PMTs arrays,
there were 143 Hamamatsu R8520-406 1-inch square PMTs at the top and
37 Hamamatsu R11410-MOD 3-inch PMTs at the bottom. After series of
engineering runs, the dark matter search data were collected between
March and October of 2014~\cite{PandaX-2014, PandaX-I-2014}. The total
exposure was $54\times80.1$~kg-day. A blind analysis was carried
out. Seven events in the signal region were found with $6.9\pm0.6$
expected background events, consistent with no excess over
background. The exclusion limit to the spin-independent (SI) isoscalar
WIMP-nucleon scattering cross section is shown in
Fig.~\ref{fig-resultpandax1}. At the $90\%$ confidence level, the
limit for WIMP mass below 5.5~GeV/$c^{2}$ was the tightest reported
constraints among all liquid xenon experiments by that time. The
exclusion limit strongly disfavored the claimed signal regions by
DAMA-LIBRA~\cite{DAMA-LIBRA2013}, CoGeNT~\cite{CoGeNT2014},
CRESST-II~\cite{CRESST-CaWO4}, and CDMS-Si~\cite{CDMS-Si}. This result
was a major milestone in the development of the PandaX program.

To improve the sensitivity to WIMPs, an immediate upgrade towards
PandaX-II was carried out in 2015. The key difference in PandaX-II was
an enlarged TPC to accommodate a 580~kg liquid xenon target. The TPC
was 60~cm high with a diameter of 60~cm. A total of 110 Hamamatsu
3-inch R11410-20 PMTs with improved quantum efficiency were
implemented. To suppress background, a new cryostat was made by
radiopure stainless steel~\cite{ZhangTaoPaper}. In the region between
the TPC field cage and the inner cryostat, two rings of 24 Hamamatsu
1-inch R8520-406 PMTs were installed at the top and bottom to further
veto external background events.

Initial dark matter data in PandaX-II were collected from November to
December in 2015 (Run 8). The run was stopped after the identification
of a high $^{85}$Kr background (a $\beta$-emitter with an abundance of
about 2$\times10^{-11}$ in natural krypton), likely introduced by an
air leak. The krypton-to-xenon atomic ratio inferred from the data was
about 400 part-per-trillion or ppt. The data taking was resumed in
March 2016 after a krypton distillation campaign, which reduced the
krypton level by about a factor of 10, leading to a record low
background rate of $2.0\times 10^{-3}~evt/(day*kg*kev)$ by that time. A low
background data set was collected between March and June 2016 (Run 9),
which, in combination with Run 8, provided a leading WIMP exposure of
33~ton-day. In the NR signal region, only one event was identified
with a mean expected background of 2.5 events. A stringent upper limit
was set on the elastic SI WIMP-nucleon scattering cross section, with
the lowest excluded cross section of $2.5\times10^{-46}$cm$^{2}$ for a
WIMP mass of 40~GeV/$c^{2}$~\cite{PandaX-II-33ton-day}, which improved
from the previous best limit from LUX~\cite{LUX-2015} by a factor of
2.5.  In addition, since the net spin of the xenon nucleus is mostly
carried the odd-neutron isotopes, $^{129}$Xe and $^{131}$Xe, the same
data provided a leading limit on the spin-dependent (SD) WIMP-neutron
interaction~\cite{PandaX-II-SD-WIMP}.

\begin{figure}[!htbp]
\centering
\includegraphics[width=0.5\textwidth]{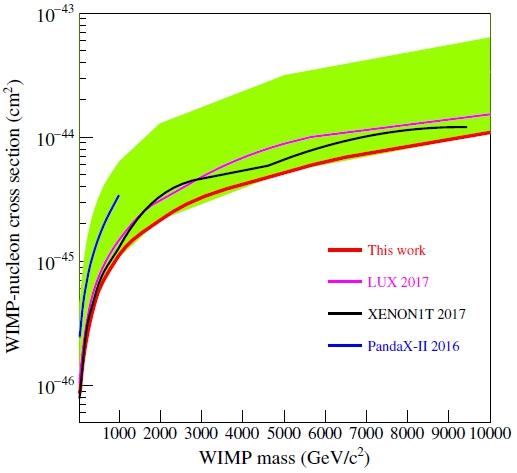}
\caption{The $90\%$ C.L. upper limits for the SI WIMP-nucleon cross
  section from the combination of the data from PandaX-II Run 9 and
  Run 10 (red solid). The 1-$\sigma$ sensitivity band is shown in
  green respectively. Data from XENON1T 2017~\cite{XENON1T-2017} and
  LUX 2017~\cite{LUX-2017} are overlaid for comparison. Figure from
  Ref.~\cite{PandaX-II-50ton-day}.  \protect\label{fig-SIpandax2-2017}}
\end{figure}

After Run 9, to make an accurate calibration of the detector response
to ER events, a tritiated methane injection was carried out, a
technique pioneered by the LUX
collaboration~\cite{LUX-CH3T-Paper}. After the calibration, to remove
the residual tritium left in the detector and to further suppress
krypton background, a second distillation campaign was carried out in
CJPL-II. The tritium was largely removed and the krypton level was
reduced to about 6~ppt. Another low background data set (Run 10) was
collected from March to July 2017. The ER background level of was
$0.8\times 10^{-3}~evt/(day*kg*kev)$, 2.5~times suppressed from that in Run
9, yet no excess was found in the signal region. The new limit to
spin-independent WIMP-nucleon interaction
(Fig.\ref{fig-SIpandax2-2017}) was released in
Ref.~\cite{PandaX-II-50ton-day} by combining Run 9 and Run 10, with a
dark matter exposure of 54~ton-day. This represented the most
stringent limit for a WIMP mass greater than 100~GeV/$c^{2}$ by that
time.

\begin{figure}[!htbp]
  \centering
  \includegraphics[width=0.48\textwidth]{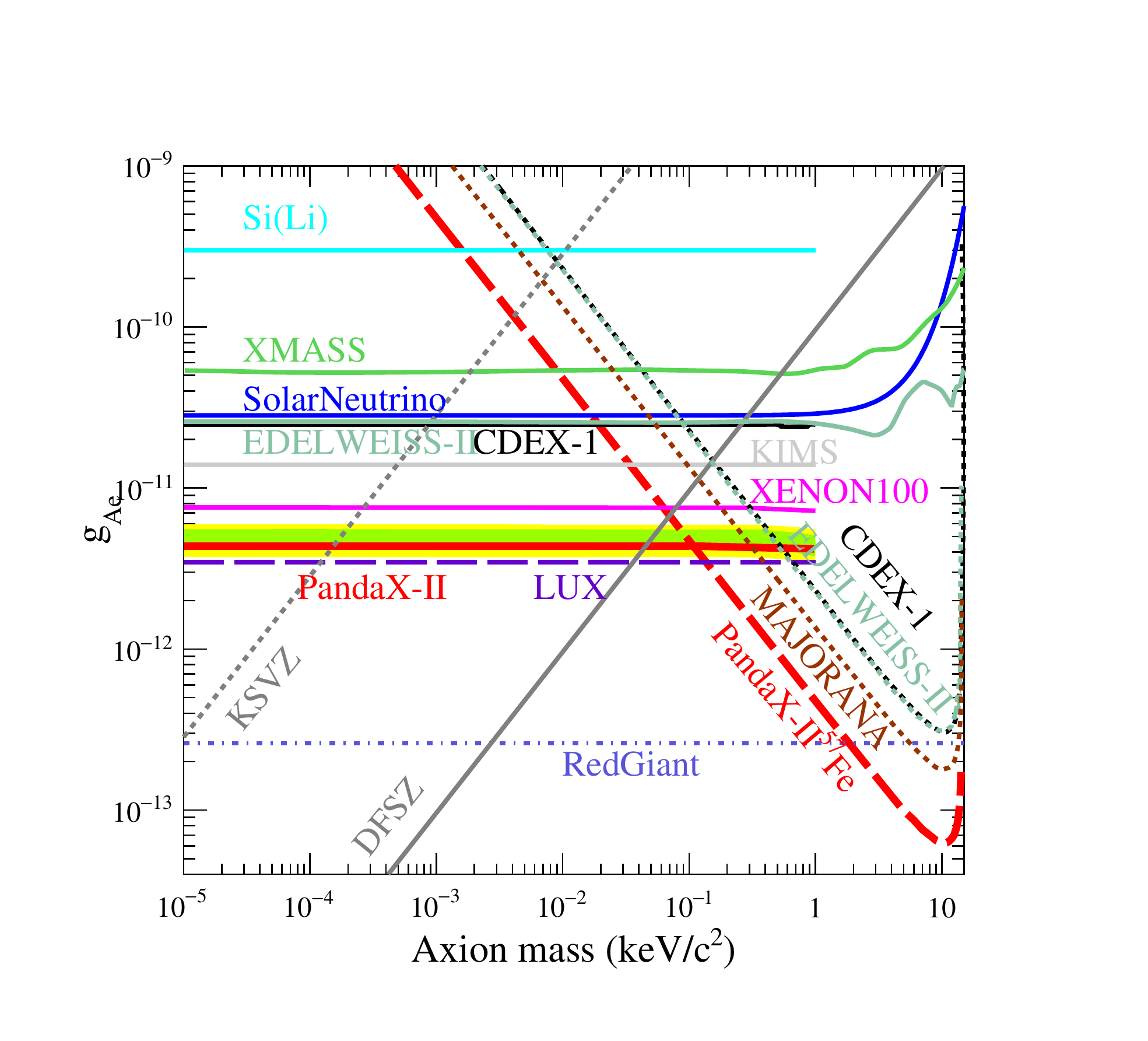}
  \includegraphics[width=0.48\textwidth]{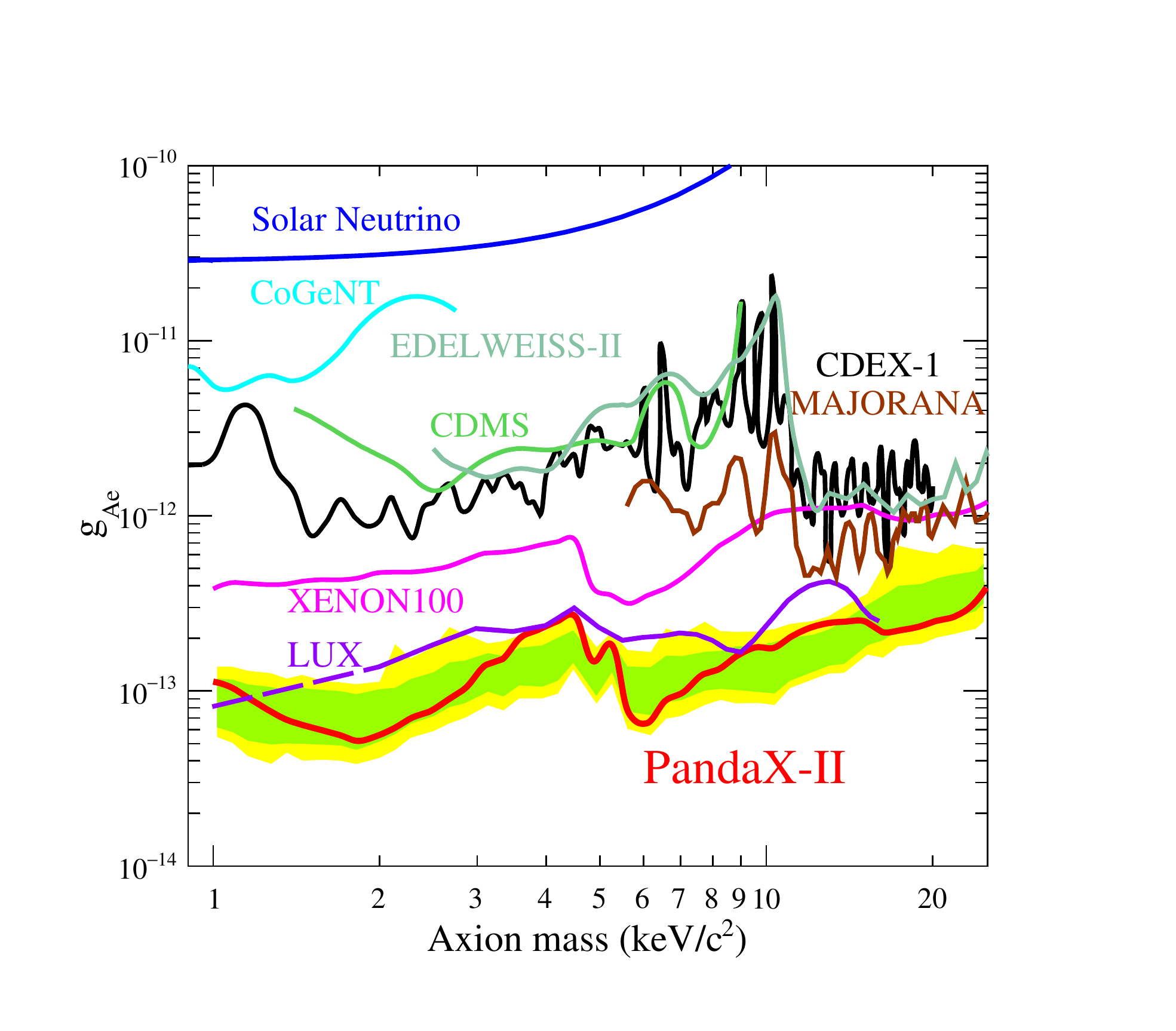}
  \caption{Axion-electron coupling constant $g_{Ae}$ vs. the axion
    mass, from PandaX-II Run 9 data, for solar axion (top) and
    galactic axion-like particles (bottom). Figure taken from
    Ref.~\cite{PandaX-II-axion}.
    \protect\label{fig-Axionpandax2-2016}}
\end{figure}

The data from PandaX-II were also used to study non-WIMP dark matter
particles and interactions. For example, axion is a pseudoscalar
particle dark matter candidate, originally proposed to mandate the
observed charge-parity or CP symmetry in the strong
interactions~\cite{Weinberg, Wilczek}. Assuming that axions can couple
to electrons, solar axion or galactic axion-like particles(ALPs) were
searched using the PandaX-II data in the ER
region~\cite{PandaX-II-axion}. Due to the low ER background rate and
large exposure, stringent limits on the dimensionless axion-electron
coupling constant $g_{Ae}$ for both scenarios were set, shown in
Fig.~\ref{fig-Axionpandax2-2016}. The PandaX-II data were also used,
the first time, to search for a special dark matter-nucleon
interactions in which the interaction portal is a light mediator
particle with a finite kinematic mixing to the known standard model
gauge particles, e.g. a photon~\cite{PandaXSIDM}. The self-interaction
between dark matters carried by such light mediator is of particular
interest to solve the so-called ``diversity problem'' in the galactic
rotational curves~\cite{HaiBoYu-PRL-2017}. For other more complex
interactions between the WIMP and nucleon beyond the standard SI and
SD interactions, a more general analysis under the effective field
theory framework is presented in Ref.~\cite{PandaXEFT2019}, in which
the state-of-art nuclear matrix element calculations for xenon nuclei
is applied. Leading constraints were set on many different forms of
interactions.

The PandaX-II operation was completed in July 2019, with an
approximate total accumulated exposure of 140 ton-day for dark matter
search. The WIMP analysis on the full data set is ongoing. Series of
detector systematics studies and technological development were also
carried out. Since natural xenon contains 8.9\% of $^{136}$Xe, a
double $\beta$-decay isotope (simultaneous conversion of two neutrons
into two protons and electrons in the nucleus), the PandaX-II
collaboration also published a new result on the neutrinoless
double-beta decay (NLDBD) search of $^{136}$Xe (with a decay Q value
of 2.458 MeV) using the data~\cite{PandaX-II-NLDBD2019}. Such a decay,
if found in nature, is a direct proof that neutrino is its own
anti-particle (so-called Majorana particle), which would also have
profound consequence in particle physics and cosmology. With the
PandaX-II data, a lower limit for the decay half-life of
2.1$\times10^{23}$ year was set at 90\% confidence level,
corresponding to an upper limit of the effective Majorana neutrino
mass between 1.4 to 3.7~eV/$c^2$. This is the first NLDBD result from a
liquid xenon dark matter experiment, which also demonstrates the
feasibility to carry out more sensitive searches in future
experiments.

The new generation of the PandaX dark matter experiment is a liquid
xenon TPC with a sensitive target of 4-ton in mass, the PandaX-4T. The
expected lowest sensitivity to WIMP-nucleon spin-independent cross
section is $6\times10^{-48}$cm$^{2}$ at 40~GeV/$c^2$ with a 6-ton-year
exposure~\cite{PandaX-4TScienceChina2019}. PandaX-4T is located in the
B2 hall of the CJPL-II. From the fall of 2019, the subsystem
components of PandaX-4T detector were transported gradually to
CJPL-II. The assembly of the detector is expected to take a full year,
then the experiment will switch to its commissioning phase. The future
generation of the PandaX experiment is also being planned as a
multi-purpose dark matter and neutrino observatory, with an ultimate
WIMP sensitivity to the ``neutrino floor''~\cite{NaturePhysics2017}.

\subsection{CDEX dark matter program}
\noindent
In 2009, the China Dark matter EXperiment (CDEX) collaboration was
established. The main goal of CDEX is to pursue studies on low mass
WIMPs ($<$10~GeV/$c^{2}$). To achieve this, it is necessary to develop
a dark matter detector with a ultra-low energy threshold. The CDEX
experiment employs the so-called P-type point-contact high purity
Germanium as the detector. Its modular structure and simple cryogenic
system make it relatively easy to scale up to large detector arrays.

\begin{figure}[!htbp]
\centering
\includegraphics[width=0.5\textwidth]{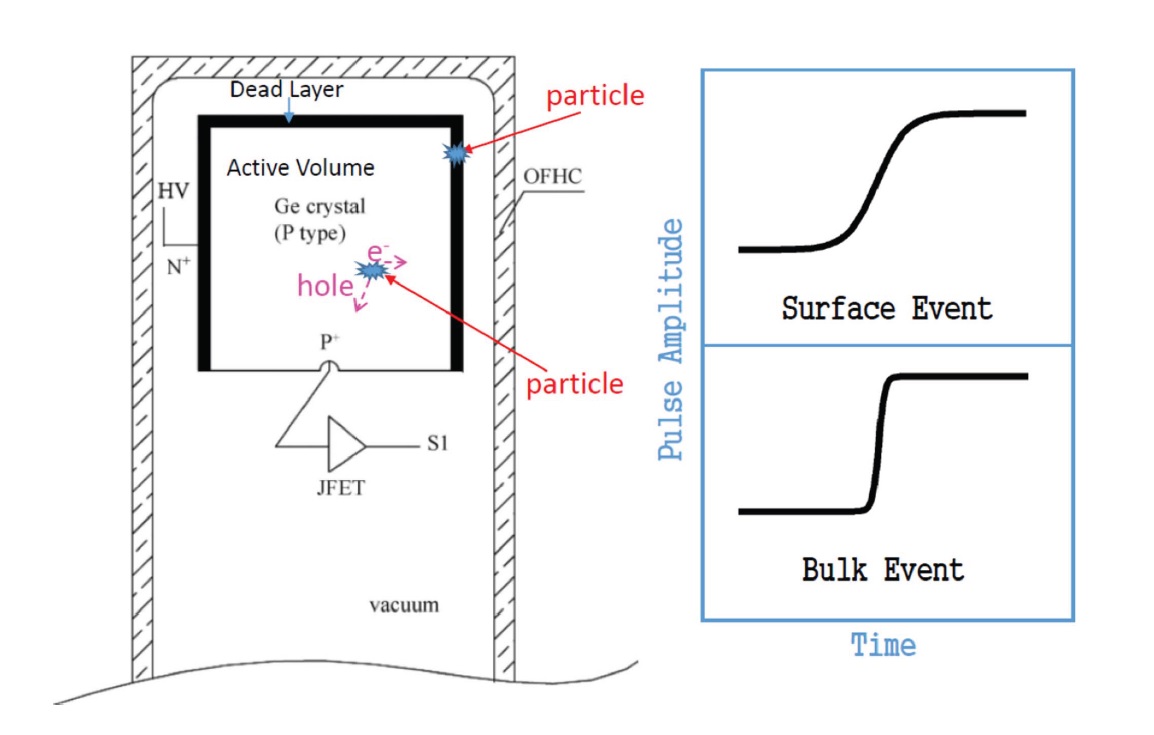}
\caption{Diagram of the PPCGe detector and sketches of the waveform
  for the surface and bulk events. Figure from
  Ref.~\cite{CDEX-zhaowei}.  \protect\label{CDEX-detector} }
\end{figure}

The point-contact technology was developed several decades ago based
on the more generally-used coaxial Germanium detector. In order to
achieve an ultra-low energy threshold, the area of the electrode is
made to be only of a mm-scale, so its capacitance can achieve the
level of few pF. This technology provides the possibility to decrease
the energy threshold down to a level of a couple of hundreds
eV. Collaborating with the Canberra Company, the CDEX collaboration
developed a P-type point-contact Germanium (PPCGe) detector with a
mass of 1~kg (CDEX-1). The structure of CDEX-1 is shown in
Fig.~\ref{CDEX-detector}. The crystal cylinder has a $n^{+}$ type
contact on the outer surface and a tiny $p^{+}$ type contact as the
central electrode. The electron-hole pairs are produced as particles
interact with Ge atoms. Under electrical field, electrons and holes
drift to the opposite electrodes. During the drift, signals will be
induced in the $p^{+}$ and $n^{+}$ electrodes. Due to the structure of
the electrode, holes close to the surface drift much slower, therefore
the signal pulse has a long rise time and a relatively small
amplitude. Therefore, such detector has the ability to differentiate
surface (background-like) and bulk (signal-like) events.

\begin{figure}[!htbp]
\centering
\includegraphics[width=0.5\textwidth]{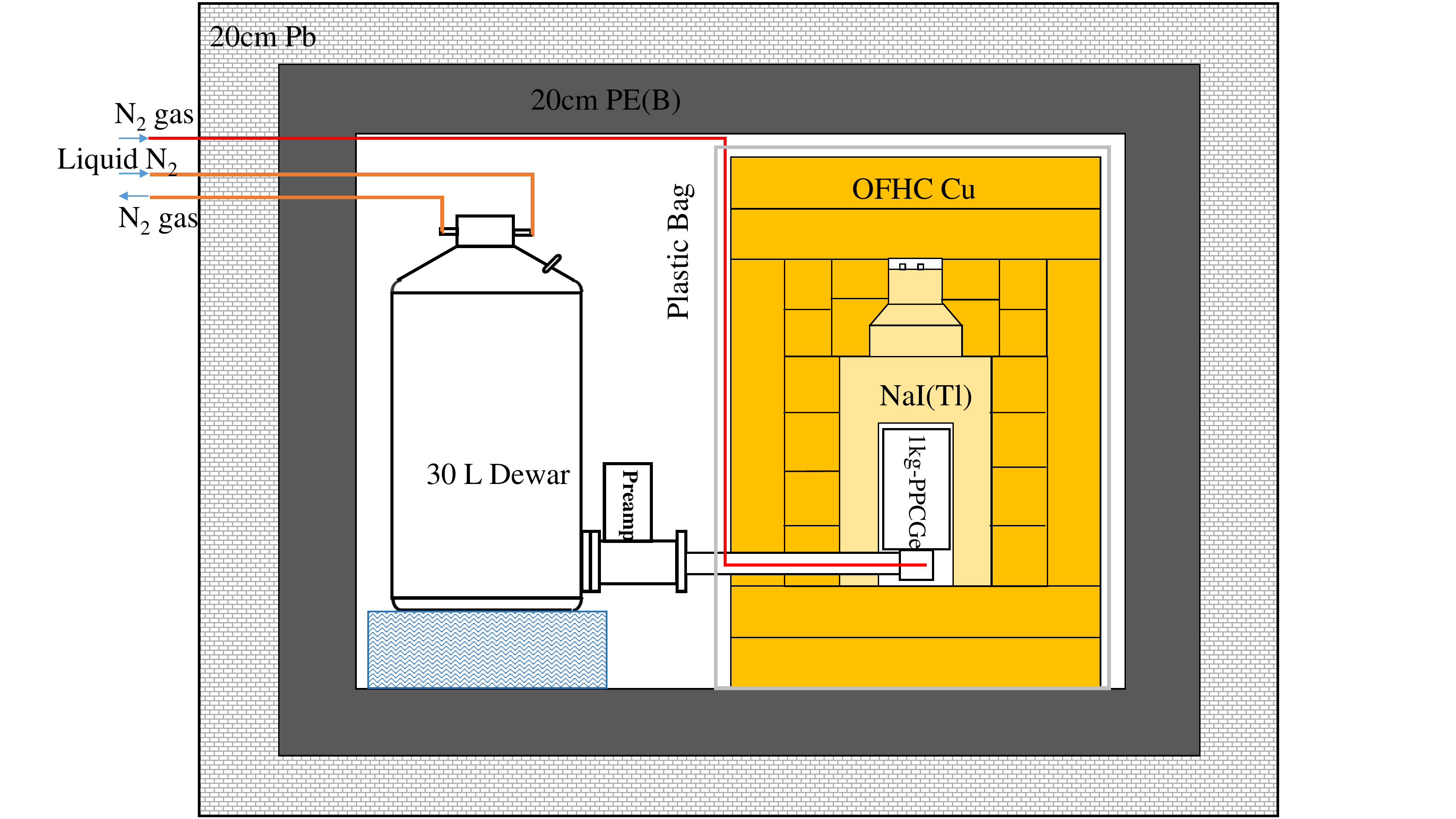}
\caption{Schematic diagram of the design of CDEX-1. The PPCGe detector
  was enclosed by an active shielding with NaI(Tl) crystal
  scintillator and a passive shielding. Figure from
  Ref.~\cite{CDEX-1-ppcGe}.  \protect\label{CDEX-shielding} }
\end{figure}

To shield the gamma rays or neutron backgrounds from the ambient
environment, a passive shielding structure has been set up for
CDEX. The outer shielding is a 1~m thick PE ``house'', containing the
entire CDEX setup. The structure of the inner shielding is shown in
Fig.~\ref{CDEX-shielding}, including a 20~cm of lead layer to stop
gamma rays and a 20~cm layer of boron-loaded PE for neutron
absorption, from outside to inside. A 20~cm thick OFHC surrounds the
cryostat of the PPCGe detector to further decrease the residual gamma
background. At the same time, the volume is purged by dry nitrogen to
suppress radon contamination. In order to further reduce the
background, an active anti-coincidence NaI(Tl) detector enclosing the
PPCGe was later implemented during the operation.

In the initial data taking period of CDEX (CDEX-1A), in order to study
the background, the bulk surface cut was not applied, nor was the
anti-coincidence detector installed. In total, 14.6~kg-days of data were
collected in 2012. This led to the first scientific result of
CDEX-1A~\cite{CDEX-1A}, in which 400~eV$_{\rm ee}$ (electron-equivalent energy) was chosen as the analysis
energy threshold. The upper limits on WIMP SI cross section at
different WIMP masses were close to the bounds from the
TEXONO~\cite{TEXONO} experiment, using the same PPCGe detector
approach. The second operation period started in 2014, with the
anti-coincidence detector implemented. CDEX-1A collected a complete
exposure of 335.6~kg-day, with a rate and spectrum above the 475~eV$_{\rm ee}$
threshold consistent with the background model. New constraints on the
SI and SD WIMP-nucleon interactions were set in
Ref.~\cite{CDEX-2014-539kgd,CDEX-1-ppcGe}, which excluded the claimed signal region
from CoGeNT using identical PPCGe technology~\cite{CoGeNT2014}. The
claimed signal region from the excess of the DAMA-LIBRA
experiment~\cite{DAMA-LIBRA2013} was also strongly disfavored.

\begin{figure}[!htbp]
\centering
\includegraphics[width=0.5\textwidth]{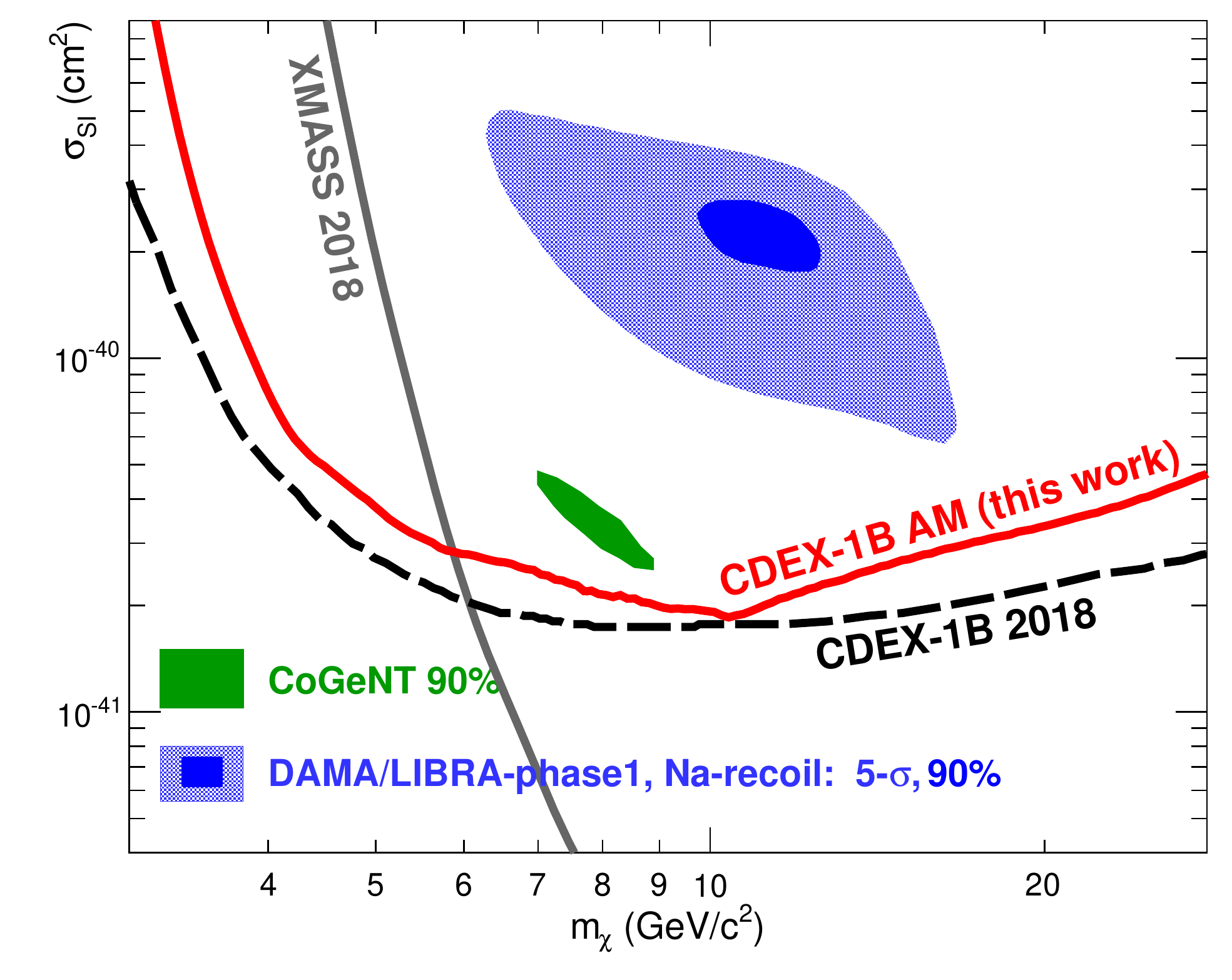}
\caption{Limits at 90\% C.L. from CDEX-1B annual modulation analysis
  (red) on spin-independent WIMP-nucleon cross section. Also shown are
  other annual modulation-based results: upper limits of XMASS-1 (dark
  gray)~\cite{XMASS-I-AM2019}, and the allowed regions from DAMA/LIBRA
  phase1~\cite{DAMA-LIBRA2013} and
  CoGeNT~\cite{CoGeNT-AM2014}. Constraints from the CDEX-1B
  time-integrated spectral analysis ~\cite{CDEX1B-TI-CPC2018} are also
  displayed (black dotted line) as comparison. Figure from
  Ref.~\cite{CDEX1B-AM-PRL2019}. \protect\label{CDEX-1B-AM2019} }
\end{figure}

To further improve the energy threshold, a new 1-kg PPCGe (CDEX-1B)
was deployed in 2014. An analysis threshold of 160 eV$_{\rm ee}$~\cite{CDEX1B-TI-CPC2018} was achieved, a
significant step forward from CDEX-1A. With a total stable data taking
span of 4.2 years, CDEX-1B was able to make an analysis on the annual
modulation of the detected event rate. Due to Earth's motion in
solar orbit, an annual modulation in the Earth's velocity relative to
the galactic DM halo thereby the collision rate with target nucleus is expected for galactic dark
matters. Both DAMA-LIBRA and the CoGeNT experiments have claimed such
evidence in their data~\cite{DAMA-LIBRA2013,CoGeNT-AM2014}. The
WIMP-nucleon SI interaction derived from the CDEX-1B annual modulation
analysis excluded the claims from DAMA-LIBRA and CoGeNT by 99.99\% and
98.0\% confidence level~\cite{CDEX1B-AM-PRL2019}
(Fig.~\ref{CDEX-1B-AM2019}).

\begin{figure}[!htbp]
\centering
\includegraphics[width=0.5\textwidth]{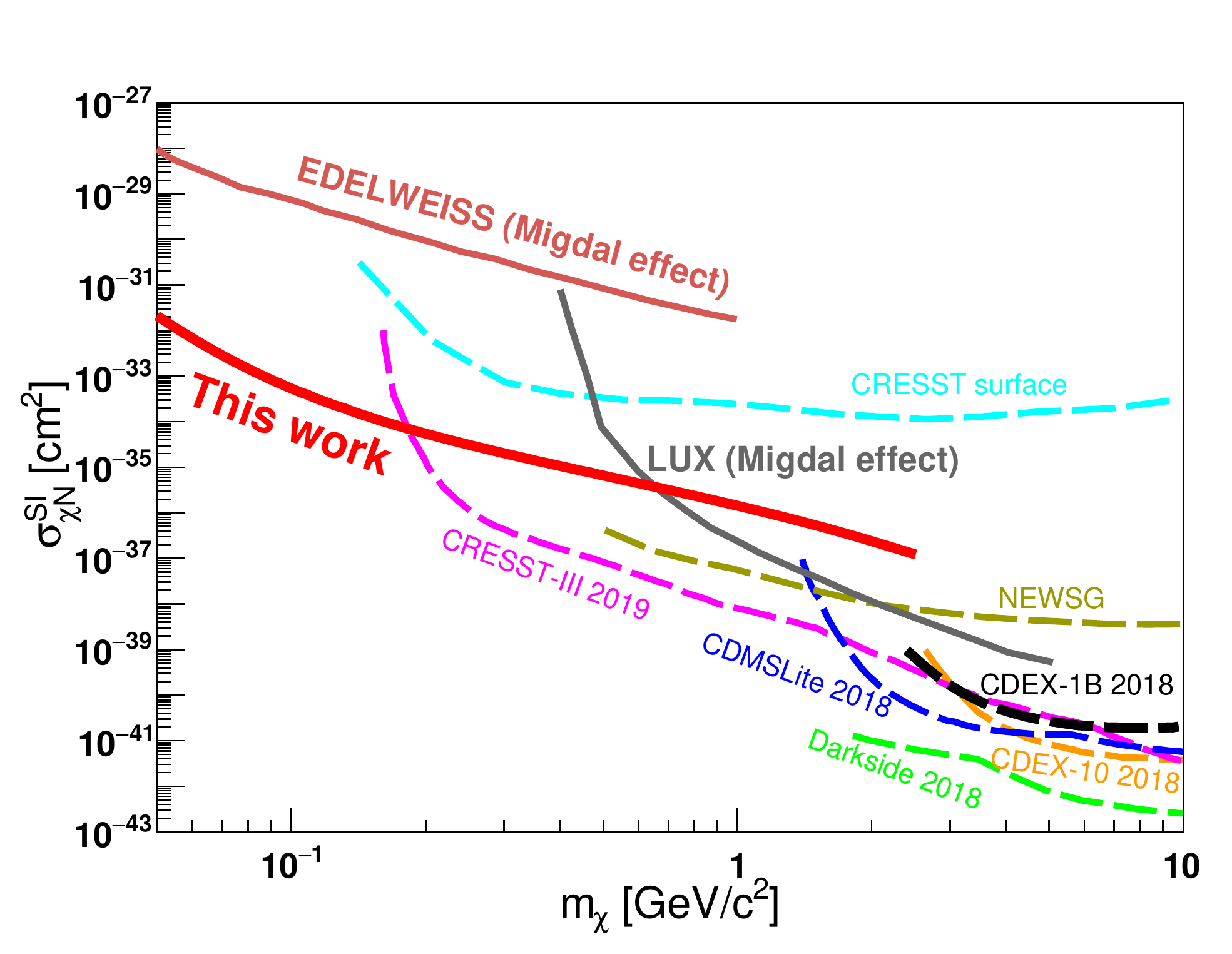}
\caption{Upper limits at 90\% C.L. on the SI dark matter-nucleon cross
  section derived by binned Poisson analysis using the CDEX-1B
  experiment data, with several benchmark
  experiments~\cite{CDMSlite-2018,DarkSide50-2018,CDEX10-2018,EDELWEISS-2019}
  superimposed. Limits from nuclear recoil-only analysis with the same
  data set is shown as well (black dash line). The limit with the
  Migdal effect incorporated (red solid line) provides the best
  sensitivities for a dark matter mass between 50 and 180
  MeV/$c^{2}$. Figure from
  Ref.~\cite{CDEX1B-SI-PRL2019}. \protect\label{CDEX-1B-ME2019} }
\end{figure}

For dark matter with lower masses (less than 1 GeV/$c^2$), the elastic
collision energy with atomic nucleus would be too small to be detected
by traditional technology. An interesting subdominant atomic effect,
the so-called Migdal effect~\cite{Migdal-Effect-DD-2018}, was recently
brought up to great attention. In short, the displacement between the
struck nucleus and the surrounding electrons could sometimes produce
additional detectable excitation energy above the detector threshold,
which in effect widens the dark matter mass range toward the lower
end. The CDEX collaboration performed this analysis
(Fig.~\ref{CDEX-1B-ME2019}), which lowered the mass range by more than
one order of magnitude (to about 50 MeV/$c^2$) in comparison to the
traditional analysis~\cite{CDEX1B-SI-PRL2019}.

\begin{figure}[!htbp]
\centering
\includegraphics[width=0.5\textwidth]{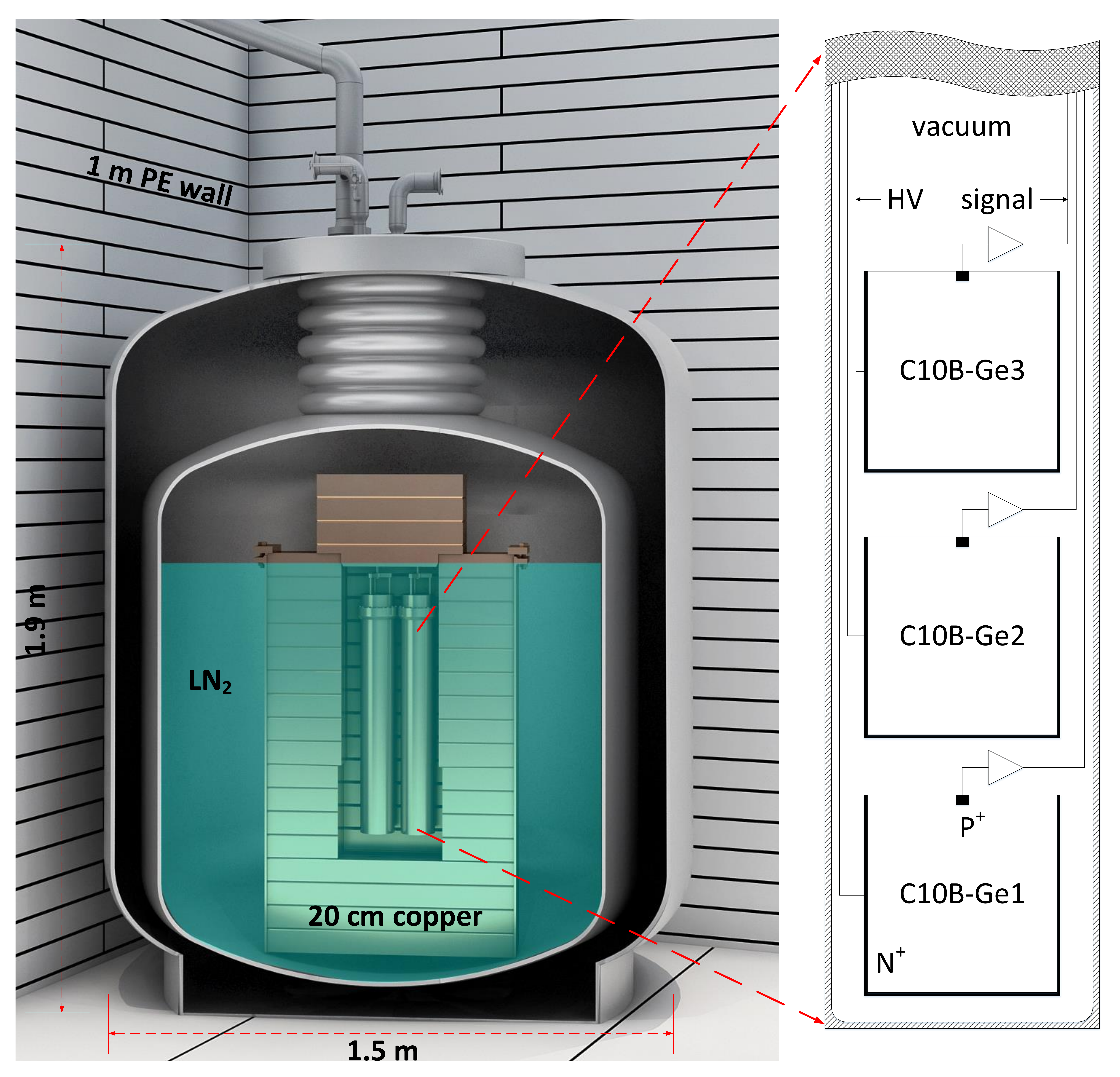}
\caption{Configuration of CDEX-10 experimental setup (left) and C10B
  detector layout in the string (right). Each detector string has
  three PPCGe detectors. Figure from Ref.~\cite{CDEX10-2018}.
  \protect\label{CDEX-10-detector}}
\end{figure}

In addition to the improvement in energy threshold, the CDEX
collaboration also made significant progress in scaling up the
detector. The upgraded CDEX experiment with a total detector mass of
about 10~kg, CDEX-10, was under operation since 2017. Three
triple-unit PPCGe strings (C10A,B,C) were directly immersed in liquid
nitrogen (see Fig.~\ref{CDEX-10-detector}). The first physics dataset
(102.8 kg-days) from one detector (C10B-Ge1) was obtained with an
analysis threshold of 160~eV$_{\rm ee}$. This results into a leading constraint
on SI WIMP-nucleon cross section at $8\times10^{-42}$~cm$^{2}$ for a
WIMP mass of 5~GeV/$c^{2}$~\cite{CDEX10-2018}.

The data from CDEX was also used to search for other rare process. For
example, CDEX-1A and CDEX-1B data provided constraints on axion-electron $g_{Ae}$ for ALPs and vector bosonic dark matter at keV-scale and below~\cite{CDEX-1A-Axion-DM2014,CDEX-1B-Axion-DM2020}. Since the
natural Germanium contains 7.8\% of $^{76}$Ge, another double
$\beta$-decay isotope, the full CDEX-1A data set was analyzed to set a
lower limit on the half-life of NLDBD of $^{76}$Ge to be
$6.4\times10^{22}$ years~\cite{CDEX-2014-onu2e}, translating into an upper
limit on the effective Majorana neutrino mass of about 5.0~eV/$c^2$.
Lately the results of dark photon searches were reported based on the CDEX-10 data,
probing new parameter space with masses from 10 to 300 eV/c2 in direct detection experiments~\cite{CDEX10-2020-darkphoton}.

The long-term goal of CDEX program is a ton-scale Germanium experiment
(CDEX-1T) searching for dark matter and NLDBD. In the C1 hall of
CJPL-II, a pit with a diameter of 18 m and a depth of 18 m was
constructed to house the future experiment. Gearing toward the future,
CDEX researchers are developing critical technologies such as
detector-grade Germanium crystal growth and ultralow background
techniques.

\section{Indirect detection in space: DAMPE satellite}
\noindent As mentioned in the introduction, two WIMP particles, when
encounter, have a finite probability to annihilate and produce the
standard model particles. Therefore, one can look for the signature of
dark matter by observing energetic $\gamma$ rays, neutrinos or charged
cosmic ray particles as the annihilation or decay products.  However,
these high energy particles could also be produced by standard
astrophysical processes. Since dark matter particle has a specific
mass, the direct annihilation products should have a narrow line feature. The
decay products or secondary particles from the annihilation would be
distributed in a particular energy range. The dark matter indirect
detection is to look for excess and features in the energy spectrum of
the detected particles, on top of the expected background spectrum
from known astrophysical processes.

The satellite-borne particle/astroparticle physics experiments started
several decades ago. The cosmic electrons, positrons, and gamma rays
have been measured by a number of satellite-borne experiments, for
example, PAMELA~\cite{DAMPE-PAMELA}, FERMI~\cite{DAMPE-FERMI},
AMS-02~\cite{DAMPE-AMS-02}, CALET~\cite{DAMPE-CALET}, and
ISS-CREAM~\cite{DAMPE-ISS-CREAM}. These experiments covered an energy
range up to 3~TeV. An intriguing broad excess in the positron to
electron ratio above a few tens GeV was found by
PAMELA~\cite{DAMPE-PAMELA-09-Nature}.  This triggers broad interests in the
community to understand both the dark matter signals as well as that
from the local astrophysical background.

The DArk Matter Particle Explorer (DAMPE) collaboration, led by the
Purple Mountain Observatory of the Chinese Academy of Sciences (CAS),
includes nine participating institutes from China, Switzerland and
Italy. DAMPE is one of the five satellite missions funded by the
Strategic Pioneer Research Program in Space Science of the CAS. The
main scientific objective of DAMPE is to measure electrons and gammas
with much higher energy resolution and energy range than previous
experiments to identify possible Dark Matter signatures.

\begin{figure}[!htbp]
\centering
\includegraphics[width=0.5\textwidth]{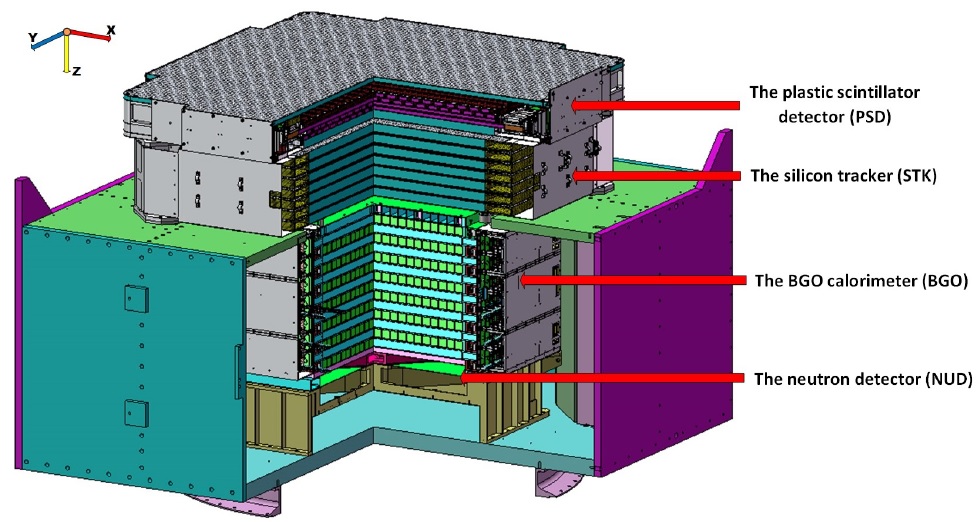}
\caption{Schematic view of the DAMPE detector. Figure from
  Ref.~\cite{DAMPE-China-detector2017}.
  \protect\label{fig-dampe-detector}}
\end{figure}

A schematics of the DAMPE detector is shown in
Fig.~\ref{fig-dampe-detector}. It consists of a Plastic Scintillator
strip Detector (PSD), a Silicon-Tungsten tracKer-converter (STK), a
BGO imaging calorimeter and a NeUtron Detector (NUD). The PSD measures
the charge of incident particles and provides charged-particle background rejection for gamma rays (anti-coincidence
detector). The STK measures the charges and the trajectories of charged particles. The
BGO calorimeter, with a total depth of about 32 radiation lengths,
allows the measurement of the energy of incident particles with high
resolution and to provide efficient electron/hadron
identification. Finally, the NUD provides an independent measurement
to further improve the electron/hadron
identification. With the combination of these four sub-detectors,
DAMPE has achieved a very effective rejection of the hadronic
cosmic-ray background, and a much improved energy resolution for
cosmic ray measurements.

In between 2014 and 2015, the DAMPE engineering qualification model
was tested using test beams at CERN. These tests demonstrated
excellent energy resolution for electrons and gamma ray (less than
1.2\% for energy $> 100$~GeV), and verified its powerful
electron/proton discrimination capabilities.

\begin{figure}[!htbp]
\centering
\includegraphics[width=0.5\textwidth]{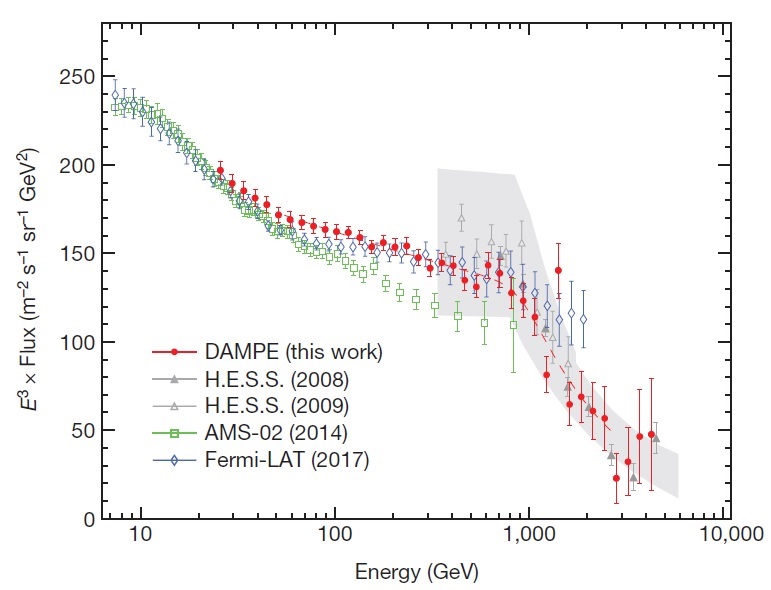}
\caption{The summed cosmic ray electrons and positrons spectrum
  ($\times E^{3}$) measured by DAMPE.The red dashed line represents a
  smoothly broken power-law model that best fits the DAMPE data in the
  range from 55 GeV to 2.63 TeV. Some results from HESS, AMS-02, and
  Fermi-LAT are overlaid for comparison (see legend). Figure from
  Ref.~\cite{DAMPE-China2017}. \protect\label{fig-dampe-result-2017}}
\end{figure}

DAMPE was launched on Dec. 17, 2015 into a sun-synchronous orbit at
the altitude of 500~km. The $e^{+}e^{-}$ analysis result from the data between 27 December 2015 and 8 June
2017 was published in 2017. The total cosmic ray $e^{+}e^{-}$ spectrum
measured by DAMPE is shown in Fig.\ref{fig-dampe-result-2017}. In the
energy range from hundreds GeV to a few TeV, these data have
unprecedented high energy resolution and low
background. Interestingly, the data can be fitted with a broken
power-law (with a break at 0.9 GeV) rather than a single power law. In
addition, a new feature was found at around 1.4 TeV,
which has triggered many theoretical discussions on the possible dark
matter origin (see, for example,
Refs.~\cite{DAMPE-thoery-YizhongFan,DAMPE-thoery-PeihongGu,DAMPE-thoery-Peter,
  DAMPE-thoery-TongLi,DAMPE-thoery-GuoliLiu,DAMPE-thoery-ShaofengGe}).
DAMPE is continuing with data taking, and the data with more
statistics are expected in the near future. Future upgrade is also
under consideration by the collaboration.

\section{Summary and perspectives}
\noindent
After almost a century since it was initially proposed, the nature of
dark matter remains mysterious. In the past decade, there is an
increasing involvement from the Chinese community in the experimental
pursue of the dark matter, with great incentive from the development
of the Jinping underground laboratory and the Chinese scientic
satellite programs. In this paper, we present the progress of the
China-led experiments in the direct and indirect dark matter
searches. The sensitivity of these highly complementary efforts reach
the forefront of the global dark matter hunt, demonstrating also the
technical capability and scientific expertise of the Chinese
community. Ambitious upgrades are also being planned for the future,
which may open up the window for a major scientific discovery.

\section{Acknowledgment}
The authors thank Prof. Qian Yue and Dr. Litao Yang from Tsinghua
  University, and also appreciate Prof. Jin Chang from Purple Mountain Observatory for
  their generous help with the content of this paper. This work is
  supported in part by the Double First Class plan of the Shanghai
  Jiao Tong University, the Key Laboratory for Particle Physics and
  Cosmology, Ministry of Education, and the Chinese Academy of
  Sciences Center for Excellence in Particle Physics. We also thank
  supports from the Natural Science Foundation of China, Ministry of
  Science and Technology, Office of Science and Technology, Shanghai
  Municipal Government, and the Hongkong Hongwen Foundation and
  Tencent Foundation in China.


~\\ ~\\
\begin{small}

\end{small}
\end{document}